\documentclass[12pt]{article}
  \usepackage{amsfonts}
  \usepackage{amsmath}
\usepackage{amssymb}
\usepackage{amscd}
\usepackage[dvips]{graphicx}

\begin{document}

~~
\bigskip
\bigskip
\begin{center}
{\Large {\bf{{{Model of hydrogen atom  for
twisted acceleration-enlarged Newton-Hooke space-times}}}}}
\end{center}
\bigskip
\bigskip
\bigskip
\begin{center}
{{\large ${\rm {Marcin\;Daszkiewicz}}$ }}
\end{center}
\bigskip
\begin{center}
{ {{{Institute of Theoretical Physics\\ University of Wroc{\l}aw pl.
Maxa Borna 9, 50-206 Wroc{\l}aw, Poland\\ e-mail:
marcin@ift.uni.wroc.pl}}}}
\end{center}
\bigskip
\bigskip
\bigskip
\bigskip
\bigskip
\bigskip
\bigskip
\bigskip
\begin{abstract}
We define the model of hydrogen atom for twist-deformed acceleration-enlarged Newton-Hooke space-time. Further, using time-dependent perturbation theory,
we find in first step of iteration procedure the solution of corresponding Schr\"{o}dinger equation as well as the probability of transition
between two different energy-eigenstates.
\end{abstract}
\bigskip
$~~~~$\;\;\;\;Pacs: 02.20.Uw, 02.40.Gh, 03.65.-w\\
\bigskip
\bigskip
\bigskip
\eject

\section{Introduction}

Recently, there appeared a lot of papers dealing with noncommutative
classical and quantum  mechanics (see e.g. \cite{mech}-\cite{qmnext})
as well as with field theoretical models (see e.g. \cite{field},
\cite{fieldnext}), in
which  the quantum space-time is employed. The suggestion to use
noncommutative coordinates goes back to Heisenberg and was firstly
formalized by Snyder in \cite{snyder}. Recently, there were also
found formal arguments based mainly on Quantum Gravity \cite{grav1} 
and String Theory models \cite{string1}, 
indicating that space-time at Planck scale  should be
noncommutative, i.e. it should  have a quantum nature. On the other
side, the main reason for such considerations follows from the
suggestion  that relativistic space-time symmetries should be
modified (deformed) at Planck scale, while  the classical Poincare
invariance still remains valid at
larger distances \cite{1a}, \cite{1anext}.

Presently, it is well known that in accordance with the
Hopf-algebraic classification of all deformations of relativistic
and nonrelativistic symmetries, one can distinguish three basic
types of quantum spaces \cite{class1}, \cite{class2}:\\
\\
{ \bf 1)} Canonical ($\theta^{\mu\nu}$-deformed) space-time
\begin{equation}
[\;{ x}_{\mu},{ x}_{\nu}\;] = i\theta_{\mu\nu}\;\;\;;\;\;\;
\theta_{\mu\nu} = {\rm const}\;, \label{noncomm}
\end{equation}
introduced in  \cite{oeckl}, \cite{chi}
in the case of Poincare quantum group and in \cite{dasz1} 
for its Galilean counterpart.\\
\\
{ \bf 2)} Lie-algebraic modification of classical space
\begin{equation}
[\;{ x}_{\mu},{ x}_{\nu}\;] = i\theta_{\mu\nu}^{\rho}{ x}_{\rho}\;,
\label{noncomm1}
\end{equation}
with  particularly chosen coefficients $\theta_{\mu\nu}^{\rho}$
being constants. This type of noncommutativity has been obtained as
the representations of the $\kappa$-Poincare \cite{kappaP} and
$\kappa$-Galilei
\cite{kappaG} as well as the  twisted relativistic \cite{lie1} 
and  nonrelativistic \cite{dasz1} 
symmetries, respectively. \\
\\
{ \bf 3)} Quadratic deformation of Minkowski and Galilei  space
\begin{equation}
[\;{ x}_{\mu},{ x}_{\nu}\;] = i\theta_{\mu\nu}^{\rho\tau}{
x}_{\rho}{ x}_{\tau}\;, \label{noncomm2}
\end{equation}
with coefficients $\theta_{\mu\nu}^{\rho\tau}$ being constants. This
kind of deformation has been proposed in \cite{qdef}, \cite{paolo},
\cite{lie1}
 at relativistic and in \cite{dasz1} at  nonrelativistic level.\\
\\
Besides, it has been demonstrated in \cite{nh} that in the case of
so-called acceleration-enlarged Newton-Hooke Hopf algebras
$\,{\mathcal U}_0(\widehat{ NH}_{\pm})$ the (twist) deformation
provides the new  space-time noncommutativity, which is expanding
($\,{\mathcal U}_0(\widehat{ NH}_{+})$) or periodic ($\,{\mathcal
U}_0(\widehat{ NH}_{-})$) in time, i.e. it takes the
form\footnote{$x_0 = ct$.}
\begin{equation}
{ \bf 4)}\;\;\;\;\;\;\;\;\;[\;t,{ x}_{i}\;] = 0\;\;\;,\;\;\; [\;{ x}_{i},{ x}_{j}\;] = 
if_{\pm}\left(\frac{t}{\tau}\right)\theta_{ij}(x)
\;, \label{nhspace}
\end{equation}
with time-dependent  functions
$$f_+\left(\frac{t}{\tau}\right) =
f\left(\sinh\left(\frac{t}{\tau}\right),\cosh\left(\frac{t}{\tau}\right)\right)\;\;\;,\;\;\;
f_-\left(\frac{t}{\tau}\right) =
f\left(\sin\left(\frac{t}{\tau}\right),\cos\left(\frac{t}{\tau}\right)\right)\;,$$
and $\theta_{ij}(x) \sim \theta_{ij} = {\rm const}$ or
$\theta_{ij}(x) \sim \theta_{ij}^{k}x_k$. Such a kind  of
noncommutativity  follows from the presence in acceleration-enlarged
Newton-Hooke symmetries $\,{\mathcal U}_0(\widehat{ NH}_{\pm})$ of
the time scale parameter (cosmological constant) $\tau$. As it was
demonstrated in \cite{nh} that  just this parameter   is responsible
for oscillation or expansion of space-time noncommutativity.

It should be noted that both Hopf structures $\,{\mathcal
U}_0(\widehat{ NH}_{\pm})$ contain, apart from rotation $(M_{ij})$,
boost $(K_{i})$ and space-time translation $(P_{i}, H)$ generators,
the additional ones denoted by $F_{i}$, responsible for constant
acceleration. Consequently, if all generators
 $F_{i}$ are equal zero we obtain the twisted Newton-Hooke quantum
 space-times \cite{nh1}, while for time parameter $\tau$ running to infinity
we get the acceleration-enlarged twisted Galilei Hopf structures proposed in
\cite{nh}. In particular,  due to the
presence of generators $F_i$, for $\tau \to \infty$  we get the new cubic and quartic type of
space-time noncommutativity
\begin{equation}
[\;{ x}_{\mu},{ x}_{\nu}\;] = i\alpha_{\mu\nu}^{\rho_1...\rho_n}{
x}_{\rho_1}...{ x}_{\rho_n}\;, \label{noncomm6}
\end{equation}
with $n=3$ and $4$ respectively, whereas for $F_i\to 0$ and $\tau \to \infty$
 we reproduce the canonical (\ref{noncomm}),
Lie-algebraic (\ref{noncomm1}) and quadratic (\ref{noncomm2})
(twisted) Galilei spaces provided in \cite{dasz1}.\\
Finally, it should be noted that all  mentioned above
noncommutative space-times have been  defined as the quantum
representation spaces, so-called Hopf modules (see
\cite{bloch}, \cite{wess}, \cite{oeckl}, \cite{chi}), for quantum acceleration-enlarged
Newton-Hooke Hopf algebras, respectively.

Recently, in the series of papers \cite{romero}-\cite{giri} there has been discussed the impact of different kinds of
quantum spaces on the dynamical structure of physical systems. Particulary, it has been demonstrated that
in the case of classical oscillator model \cite{oscylator} as well as in the case of nonrelativistic particle moving in constant
external field force $\vec{F}$ \cite{daszwal}, there are generated by space-time noncommutativity additional force terms. Such a type
of investigation has been  performed for quantum oscillator model as well \cite{oscylator}, i.e. it was demonstrated that the  quantum space in nontrivial way affects on the spectrum of energy operator. Besides, in article \cite{toporzelek} there has been considered model of particle moving on the $\kappa$-Galilei space-time in the presence of gravitational field force. It has been demonstrated that in such a case
there is produced force term, which can be identified with so-called Pioneer anomaly \cite{pioneer}, and the value of deformation parameter $\kappa$ can be fixed by comparison of obtained result with observational data.

From the abovementioned point of view, the case of two spatial directions commuting to the function of classical time seems to be most
interesting. As it was demonstrated in articles \cite{oscylator} and \cite{daszwal}, just this type of space-time noncommutativity produces additional time-dependent force terms, which appear in Hamiltonian function of the models. Usually, such a situation is interpreted as the interaction (by  radiation)  of considered system with (some) external source \cite{24}, \cite{24a}.

In this article we consider the hydrogen atom model defined on the twist-deformed acceleration-enlarged Newton-Hooke space-times (\ref{spaces})\footnote{The choice of acceleration-enlarged Newton-Hooke quantum spaces is dictated by the fact that they constitute the most general known deformation of classical nonrelativistic space-time.}. As one can see, it introduces due to the presence of function $f(t)$ the time-dependent interaction term $V_{\rm{int}}(t,\bar{x})$\footnote{Particulary, for function $f(t) = f_{\kappa_1}(t) = \kappa_1$ we recover the well-known
noncommutative hydrogen atom model proposed in paper \cite{qmnext}.}. Such a system can be analyzed with use of so-called time-dependent perturbation theory \cite{24}, \cite{24a}, i.e. one can find by iteration procedure solution of the corresponding Schr\"{o}dinger equation as well as the probability of transition
between two different energy-eigenstates. Surprisingly, for  particular choice of function $f(t)$, we get the interaction term similar to the potential function describing hydrogen atom present  in time-periodic  electric field \cite{24a}. Such a result indicates that the choice of different functions $f(t)$ may correspond to the choice of different kinds of external radiation  sources.

The paper is organized as follows. In first section we recall commonly-known facts which  concern hydrogen atom model defined on classical  space-time. In Section 2 we introduce its noncommutative  counterpart given on twist-deformed acceleration-enlarged Newton-Hooke space-time. Particulary, we find the solution of corresponding Schr\"{o}dinger equation at first order of perturbation series as well as we calculate the probability of transition between two different energy-eigenstates. The final remarks are discussed in the last section.

\section{Model of hydrogen atom  in commutative space-time}

In this section we recall basic facts associated with hydrogen atom model defined on commutative space-time \cite{24},
\cite{24a}. Firstly, it should be noted that the dynamics of considered system  is described by the following Schr\"{o}dinger equation
\begin{equation}
i\hbar \frac{\partial \psi (t,\bar{x})}{\partial t} = H(\bar{p},\bar{x})\psi (t,\bar{x})\;,\label{equation1}
\end{equation}
where
\begin{eqnarray}
H(\bar{p},\bar{x}) = H_{0}(\bar{p}) + V(\bar{x})\;\;\;;\;\;\;H_{0}(\bar{p}) = \frac{\bar{p}^2}{2m}\;\;\;,\;\;\;V(\bar{x})=-\frac{Ze^2}{r}\;,\label{equation2}
\end{eqnarray}
and
\begin{eqnarray}
r=\sqrt{x_1^2+x_2^2+x_3^2}\;\;\;,\;\;\;\bar{a}=(a_1,a_2,a_3)\;\;\;,\;\;\;Z=1\;.\label{equation3}
\end{eqnarray}
It is well-known that the energy spectrum of such defined model is degenerate, i.e. the eigenvalues of the energy operator (\ref{equation2}) take the form
\begin{equation}
E_{n} = -\frac{me^4Z^2}{2\hbar n^2}\;\;\;;\;\;\;n=1,2,3, \ldots\;,\label{equation4}
\end{equation}
while the corresponding eigenfunctions look as follows
\begin{equation}
\psi_{nlm}(r,\theta,\varphi) = N_{nlm}{\rm e}^{-\frac{r}{r_0}}\left(\frac{r}{r_0}\right)^lL_{n+l}^{(2l+1)}\left(2\frac{r}{r_0}\right)
P_l^m\left({\rm cos}\theta\right) {\rm e}^{im\varphi}
\;,\label{equation5}
\end{equation}
with $l\leq n-1$, $-l<m<l$,
\begin{eqnarray}
N_{nlm} &=& N_{nl}N_{lm}N_{m}\;\;\;,\;\;\;N_{m} \;=\; \frac{1}{2\pi}\;\;\;,\;\;\;a \;=\;\frac{\hbar^2}{me^2}\;\;\;,\;\;\;r_{0}
\;=\;\sqrt{\frac{-\hbar^2}{2mE}}
\;,\label{equation6}\\
N_{lm} &=& \sqrt{\frac{(2l+1)(l-m)!}{2(l+m)!}}\;\;\;,\;\;\;N_{nl}\;=\;
\left(\frac{2Z}{a}\right)^{3/2}\frac{1}{n^2}\sqrt{\frac{(n-l-1)!}{2[(n+l)]^3!}}\;,
\label{equation7}
\end{eqnarray}
and $L_{n}(x)$, $P^m_n(x)$ being the Laguere and Legendre polynomials respectively. \\
Besides, one can  check that the above eigenvectors constitute the orthonormal base in Hilbert space of square integrated functions $L^2(\mathbb{R}^3;d^3x)$
\begin{equation}
\left(\psi_{nlm}(\bar{x}),\psi_{n'l'm'}(\bar{x})\right) = \delta_{nn'}\delta_{mm'}\delta_{ll'}\;.\label{equation8}
\end{equation}
Finally, it should be noted that due to the stationary character of potential function (\ref{equation3}), the solution of Schr\"{o}dinger equation (\ref{equation1}) is given by
\begin{equation}
\psi(t,\bar{x}) = \sum_{n=0}^{\infty}\sum_{l=0}^{n-1}\sum_{m=-l}^{l} c_{nlm}{\rm e}^{iE_n(t-t_0)}\psi_{nlm}(\bar{x}) \;,\label{equation9}
\end{equation}
with symbol $t_0$ denoting initial time of evolution. \\
\\
\\

\section{Model of hydrogen atom  for twisted acceleration-enlarged Newton-Hooke space-times}

\subsection{Twisted acceleration-enlarged Newton-Hooke space-times}

In this subsection we turn to the twisted acceleration-enlarged Newton-Hooke space-times  equipped with two spatial directions
commuting to classical time, i.e. we consider  spaces of the form \cite{nh}
\begin{equation}
[\;t,\hat{x}_{i}\;] =[\;\hat{x}_{1},\hat{x}_{3}\;] = [\;\hat{x}_{2},\hat{x}_{3}\;] =
0\;\;\;,\;\;\; [\;\hat{x}_{1},\hat{x}_{2}\;] =
if({t})\;\;;\;\;i=1,2,3
\;, \label{spaces}
\end{equation}
with  function $f({t})$ given by\footnote{The indecies $+$ and $-$ in formulas (\ref{w2})-(\ref{w7}) correspond to the acceleration-enlarged Newton-Hooke quantum spaces associated with Hopf structures, which can be get partially by the contraction of De-Sitter and Anti-De-Sitter Hopf algebras respectively.}
\begin{eqnarray}
f({t})&=&f_{\kappa_1}({t}) =
f_{\pm,\kappa_1}\left(\frac{t}{\tau}\right) = \kappa_1\,C_{\pm}^2
\left(\frac{t}{\tau}\right)\;, \label{w2}\\
f({t})&=&f_{\kappa_2}({t}) =
f_{\pm,\kappa_2}\left(\frac{t}{\tau}\right) =\kappa_2\tau\, C_{\pm}
\left(\frac{t}{\tau}\right)S_{\pm} \left(\frac{t}{\tau}\right) \;,
\label{w3}\\
f({t})&=&f_{\kappa_3}({t}) =
f_{\pm,\kappa_3}\left(\frac{t}{\tau}\right) =\kappa_3\tau^2\,
S_{\pm}^2 \left(\frac{t}{\tau}\right) \;, \label{w4}\\
f({t})&=&f_{\kappa_4}({t}) =
 f_{\pm,\kappa_4}\left(\frac{t}{\tau}\right) = 4\kappa_4
 \tau^4\left(C_{\pm}\left(\frac{t}{\tau}\right)
-1\right)^2 \;, \label{w5}\\
f({t})&=&f_{\kappa_5}({t}) =
f_{\pm,\kappa_5}\left(\frac{t}{\tau}\right) = \pm \kappa_5\tau^2
\left(C_{\pm}\left(\frac{t}{\tau}\right)
-1\right)C_{\pm} \left(\frac{t}{\tau}\right)\;, \label{w6}\\
f({t})&=&f_{\kappa_6}({t}) =
f_{\pm,\kappa_6}\left(\frac{t}{\tau}\right) = \pm \kappa_6\tau^3
\left(C_{\pm}\left(\frac{t}{\tau}\right) -1\right)S_{\pm}
\left(\frac{t}{\tau}\right)\;; \label{w7}
\end{eqnarray}
$$C_{+/-} \left(\frac{t}{\tau}\right) = \cosh/\cos \left(\frac{t}{\tau}\right)\;\;\;{\rm and}\;\;\;
S_{+/-} \left(\frac{t}{\tau}\right) = \sinh/\sin
\left(\frac{t}{\tau}\right) \;.$$
As it was already mentioned in Introduction, in $\tau \to \infty$ limit the above quantum spaces reproduce the canonical (\ref{noncomm}),
Lie-algebraic (\ref{noncomm1}), quadratic (\ref{noncomm2}) as  well as cubic and quartic (\ref{noncomm6}) type of
space-time noncommutativity, with\footnote{Space-times (\ref{nw2}), (\ref{nw4}) correspond to the
twisted Galilei Hopf algebras provided in \cite{dasz1}, while the quantum space (\ref{nw7}) is
associated with acceleration-enlarged Galilei Hopf structure \cite{nh}.},\footnote{Technically, we expand the r.h.s of formulas (\ref{w2})-(\ref{w7})
in Taylor series with respect variable $\frac{t}{\tau}$ and take the limit $\tau \to \infty$. In such a way there survive only in the commutation
relation (\ref{nw2})-(\ref{nw7}) $t^n$ terms with $n = 0,1,2,3$ and $4$ respectively.} 
\begin{eqnarray}
f_{\kappa_1}({t}) &=& \kappa_1\;,\label{nw2}\\
f_{\kappa_2}({t}) &=& \kappa_2\,t\;,\label{nw3}\\
f_{\kappa_3}({t}) &=& \kappa_2\,t^3\;,\label{nw4}\\
f_{\kappa_4}({t}) &=& \kappa_4\,t^4\;,  \label{nw5}\\
f_{\kappa_5}({t}) &=& \frac{1}{2}\kappa_5\,t^2\;, \label{nw6}\\
f_{\kappa_6}({t}) &=& \frac{1}{2}\kappa_6\,t^3\;. \label{nw7}
\end{eqnarray}
Of course, for all  parameters $\kappa_a$ $(a=1,...,6)$ running to zero the above deformations disappear.

\subsection{Model of hydrogen atom}

Let us now turn to the main aim of our investigations - to the model of hydrogen atom defined on quantum space-times (\ref{w2})-(\ref{w7}).
In first step of our construction we extend the described in pervious subsection spaces to the whole algebra of momentum and position operators as follows
\begin{eqnarray}
&&[\;\hat{ x}_{1},\hat{ x}_{2}\;] = if_{\kappa_a}({t})\;\;\;,\;\;\;[\;\hat{ x}_{1},\hat{ x}_{3}\;] =
 [\;\hat{ x}_{2},\hat{ x}_{3}\;] =
 [\;\hat{ p}_{i},\hat{ p}_{j}\;] =0\;,\label{phasespaces1}\\
&&~~~~~~~~~~~~~~~~~[\;\hat{ x}_{i},\hat{ p}_{j}\;] = {i\hbar}\delta_{ij}\;\;;\;\;i,j=1,2,3\;. \label{phasespaces2}
\end{eqnarray}
One can check that relations (\ref{phasespaces1}), (\ref{phasespaces2}) satisfy the Jacobi identity and for deformation parameters
$\kappa_a$ approaching zero become classical. \\
Next, by analogy to the commutative case we define the Hamiltonian operator
\begin{eqnarray}
H(\hat{p},\hat{x}) = \frac{\bar{\hat{p}}^2}{2m} -\frac{Ze^2}{\hat{r}}\;,\label{grom1}
\end{eqnarray}
with $\hat{r}\;=\;\sqrt{\hat{x}_1^2+\hat{x}_2^2+\hat{x}_3^2}$. \\
In order to analyze the above system we represent the
noncommutative operators $({\hat x}_i, {\hat p}_i)$ by classical
ones $({ x}_i, { p}_i)$ as  (see e.g.
\cite{kijanka}-\cite{ggggg})
\begin{equation}
{\hat x}_{1} = { x}_{1} - \frac{f_{\kappa_a}(t)}{2\hbar}
p_2\;\;\;,\;\;\;{\hat x}_{2} = { x}_{2} +\frac{f_{\kappa_a}(t)}{2\hbar}
p_1\;\;\;,\;\;\; {\hat x}_{3}= x_3 \;\;\;,\;\;\; {\hat p}_{i}=
p_i\;, \label{rep}
\end{equation}
where
\begin{equation}
[\;x_i,x_j\;] = 0 =[\;p_i,p_j\;]\;\;\;,\;\;\; [\;x_i,p_j\;]
={i\hbar}\delta_{ij}\;. \label{classpoisson}
\end{equation}
Then, the  Hamiltonian (\ref{grom1}) takes the form
\begin{eqnarray}
H(\bar{p},\bar{x},t) &=& \frac{\bar{p}^2}{2m} -\frac{Ze^2}{{r}} - \frac{Ze^2L_3f_{\kappa_a}(t)}{2\hbar{r^3}} + \mathcal{O}
(\kappa_a) \;= ~~~~~~~~~~~~~~~~~~~~~~\label{grom2}\\
&~~&~~~~~~~~~~~~~~~~~~~~~~=\;H_{0}(\bar{p},\bar{x}) + V_{\rm{int}}(t,\bar{x})+ \mathcal{O}
(\kappa_a)\;,\nonumber
\end{eqnarray}
with $\mathcal{O}(\kappa_a)$ denoting terms quadratic in deformation parameter $\kappa_a$ and $L_3 =x_1p_2 - x_2p_1$. Besides, present in the above formula additional
 potential function
\begin{eqnarray}
V_{\rm{int}}(t,\bar{x}) = - \frac{Ze^2L_3f_{\kappa_a}(t)}{2\hbar{r^3}}\;,\label{grom3}
\end{eqnarray}
describes interaction (by radiation) of quantum particle with (some) external source.

It is well-known that such a system can be analyzed with use of time-dependent perturbation theory \cite{24}, \cite{24a}. Then, the
solution of corresponding Schr\"{o}dinger equation
\begin{equation}
i\hbar \frac{\partial \psi (t,\bar{x})}{\partial t} = \left[\;H_0(\bar{p},\bar{x})+V_{\rm{int}}(t,\hat{x})\;\right]\psi (t,\bar{x})\;,\label{grom4}
\end{equation}
is given by
\begin{equation}
\psi(t,\bar{x}) = \sum_{n=0}^{\infty}\sum_{l=0}^{n-1}\sum_{m=-l}^{l} c_{nlm}(t){\rm e}^{iE_n(t-t_0)}\psi_{nlm}(\bar{x}) \;,\label{grom5}
\end{equation}
where  symbols $E_n$ and $\psi_{nlm}(\bar{x})$ denote eigenvalues (\ref{equation4}) and eigenfunctions (\ref{equation5}) for classical hydrogen atom respectively.  The coefficients $c_{nlm}(t)$ are defined as the solutions of the following differential equations
\begin{eqnarray}
\frac{dc_{nlm}(t)}{dt} &=& \frac{1}{i\hbar} \sum_{n'=0}^{\infty}\sum_{l'=0}^{n-1}\sum_{m'=-l}^{l} \left(\psi_{nlm}(\bar{x}),V_{\rm{int}}(t,\hat{x})\psi_{n'l'm'}(\bar{x})\right)c_{n'l'm'}(t_0)\;\cdot \cr
&\cdot&{\rm e}^{i\omega_{nn'}(t-t_0)}\;\;\;;\;\;\;\omega_{nn'} \;=\; \frac{1}{\hbar}(E_n-E_{n'})\;,\label{grom6}
\end{eqnarray}
with initial condition of the form $c_{n'l'm'}(t_0) = \left(\psi_{n'l'm'}(\bar{x}),\psi_0(\bar{x})\right)$ for some wave function $\psi_{0}(\bar{x})$. Usually, the above system can be solve by iteration procedure with zero-approximated term given by
\begin{eqnarray}
c_{nlm}^{(0)}(t) = \left(\psi_{nlm}(\bar{x}),\psi_0(\bar{x})\right)\;,\label{grom7}
\end{eqnarray}
and with two neighboring higher steps of approximation combined into the following equation
\begin{eqnarray}
\frac{dc_{nlm}^{(j)}(t)}{dt}
&=& \frac{1}{i\hbar} \sum_{n'=0}^{\infty}\sum_{l'=0}^{n-1}\sum_{m'=-l}^{l} \left(\psi_{nlm}(\bar{x}),V_{\rm{int}}(t,\bar{x})\psi_{n'l'm'}(\bar{x})\right)c_{n'l'm'}^{(j-1)}(t)\;\cdot \cr
&\cdot&{\rm e}^{i\omega_{nn'}(t-t_0)}
\;.\label{grom8}
\end{eqnarray}
Particulary, in accordance with the above formula, for $j=1$ and for $\psi_0(\bar{x})=\psi_{n''l''m''}(\bar{x})$, we get ($t_0=0$)
\begin{eqnarray}
c_{nlm}^{(1)}(t) = \delta_{nn''}\delta_{ll''}\delta_{mm''} + \frac{1}{i\hbar} \int_{0}^{t}dt_1{\rm e}^{i\omega_{nn''}t_1}
\left(\psi_{nlm}(\bar{x}),V_{\rm{int}}(t_1,\bar{x})\psi_{n''l''m''}(\bar{x})\right)\;,\label{grom9}
\end{eqnarray}
and, by direct calculation, we obtain
\begin{eqnarray}
c_{nlm}^{(1)}(t) &=& \delta_{ll''}\delta_{mm''} - \frac{m''Ze^2}{i2\hbar^2} g(t)
\left(\psi_{nlm}(\bar{x}),\frac{1}{r^3}\psi_{n''l''m''}(\bar{x})\right) \;;\label{grom10}\\
g({t})&=&g_{\kappa_1}({t}) =
g_{\pm,\kappa_1}\left(\frac{t}{\tau}\right) = \frac{\kappa_1}{4}\left( \tau S_{\pm}
\left(\frac{2t}{\tau}\right) +2t \right)\;, \label{gromw2}\\
g({t})&=&g_{\kappa_2}({t}) =
g_{\pm,\kappa_2}\left(\frac{t}{\tau}\right) =\pm \frac{\kappa_2}{4}\tau^2\left( C_{\pm}
\left(\frac{2t}{\tau}\right)-1\right) \;,
\label{gromw3}\\
g({t})&=&g_{\kappa_3}({t}) =
g_{\pm,\kappa_3}\left(\frac{t}{\tau}\right) =\pm\frac{\kappa_3}{4}\tau^2\left(\tau
S_{\pm} \left(\frac{2t}{\tau}\right)-2t\right) \;, \label{gromw4}\\
g({t})&=&g_{\kappa_4}({t}) =
 g_{\pm,\kappa_4}\left(\frac{t}{\tau}\right) = \kappa_4
 \tau^4\left(\tau S_{\pm}\left(\frac{2t}{\tau}\right)
-8\tau S_{\pm}\left(\frac{t}{\tau}\right) +6t\right) \;, \label{gromw5}\\
g({t})&=&g_{\kappa_5}({t}) =
g_{\pm,\kappa_5}\left(\frac{t}{\tau}\right) = \pm \frac{\kappa_5}{4}\tau^2
\left(\tau S_{\pm}\left(\frac{2t}{\tau}\right)
-4\tau S_{\pm} \left(\frac{t}{\tau}\right) + 2t\right)\;, \label{gromw6}\\
g({t})&=&g_{\kappa_6}({t}) =
g_{\pm,\kappa_6}\left(\frac{t}{\tau}\right) =  \frac{\kappa_6}{2}\tau^4
\left(\left(C_{\pm}\left(\frac{t}{\tau}\right)-2\right)C_{\pm}\left(\frac{t}{\tau}\right) +1\right)\;, \label{gromw7}
\end{eqnarray}
with $n=n''$ and
\begin{eqnarray}
c_{nlm}^{(1)}(t) &=& -\frac{m''Ze^2}{i2\hbar^2} h(t)
\left(\psi_{nlm}(\bar{x}),\frac{1}{r^3}\psi_{n''l''m''}(\bar{x})\right) \;;\label{grom10a}\\
h({t})&=&h_{\kappa_1}({t}) =
h_{\pm,\kappa_1}\left(\frac{t}{\tau}\right) = -\frac{i\kappa_1}{2 \omega_{nn''} (\tau^2 \omega_{nn''}^2 \pm 4)}\left[ {\rm e}^{i t \omega_{nn''}} \left(\tau^2 \omega_{nn''}^2 C_{\pm}\left(\frac{2 t}{\tau}\right)+\right.\right.\nonumber \\
&+&\left.\left.\tau^2 \omega_{nn''}^2\pm 2 i \tau \omega_{nn''}S_{\pm}\left(\frac{2 t}{\tau}\right)\pm 4\right)-2\tau^2 \omega_{nn''}^2 \mp 4\right]\;, \label{ngromw2}\\
h({t})&=&h_{\kappa_2}({t}) =
h_{\pm,\kappa_2}\left(\frac{t}{\tau}\right) =
\frac{\kappa_2\tau^2 }{2 (\tau^2 \omega_{nn''}^2 \pm 4)}
 \left[{\rm e}^{i t \omega_{nn''}}\left(2 C_{\pm}\left(\frac{2 t}{\tau}\right)-i \tau \omega_{nn''}\right.\right.\cdot \nonumber\\
  &\cdot& \left.\left.S_{\pm}\left(\frac{2 t}{\tau}\right)\right)-2\right] \;,
\label{ngromw3}\\
h({t})&=&h_{\kappa_3}({t}) =
h_{\pm,\kappa_3}\left(\frac{t}{\tau}\right) =\mp \frac{i\kappa_3\tau^2}{2 \omega_{nn''} (\tau^2 \omega_{nn''}^2\pm 4)} \left[{\rm e}^{i t \omega_{nn''}}
\left(\tau^2 \omega_{nn''}^2 C_{\pm}\left(\frac{2 t}{\tau}\right)+\right.\right.\nonumber\\
&-&\left.\left.\tau^2 \omega_{nn''}^2\pm 2 i \tau \omega_{nn''}S_{\pm}\left(\frac{2 t}{\tau}\right)\mp 4\right)\pm 4
\right]
\;, \label{ngromw4}\\
h({t})&=&h_{\kappa_4}({t}) =
 h_{-,\kappa_4}\left(\frac{t}{\tau}\right) =
 \frac{\kappa_4}{2} \tau^4{\rm e}^{i t \omega_{nn''}}\left(C_{-}\left(\frac{t}{\tau}\right)-1\right)^2
\csc^4\left(\frac{t}{2\tau}\right)\cdot\nonumber\\
&\cdot&\left[\frac{4\tau S_{-}\left(\frac{t}{\tau}\right)}{\tau^2\omega_{nn''}^2-1} -
\frac{2\tau S_{-}\left(\frac{2t}{\tau}\right)}{\tau^2\omega_{nn''}^2-4} + \frac{4i\tau^2 \omega_{nn''}
C_{-}\left(\frac{t}{\tau}\right)}{\tau^2\omega_{nn''}^2-1} - \frac{i\tau^2\omega_{nn''}C_{-}\left(\frac{2t}{\tau}\right)} {\tau^2\omega_{nn''}^2-4} +\right. \nonumber\\
&-&\left.\frac{3i}{\omega_{nn''}}\right]+ \frac{24i\kappa_4\tau^2}{\tau^4\omega_{nn''}^5-5\tau^2\omega^3+4\omega_{nn''}}
  \;, \label{ngromw5}\\
 h({t})&=&h_{\kappa_4}({t}) =
 h_{+,\kappa_4}\left(\frac{t}{\tau}\right) = -\frac{2i\kappa_4\tau^4}{\omega_{nn''}(\tau^4\omega_{nn''}^4+5\tau^2
 \omega_{nn''}^2+4)}\left[{\rm e}^{i t \omega_{nn''}}\left(3\tau^4\omega_{nn''}^4 +\right.\right.\nonumber\\
 &-&\left. 4i\tau^3\omega_{nn''}^3S_{+}\left(\frac{t}{\tau}\right)+ 2i\tau^3\omega_{nn''}^3S_{+}\left(\frac{2t}{\tau}\right)
 -4\tau^2\omega_{nn''}^2 (\tau^2\omega_{nn''}^2 + 4)C_{+}\left(\frac{t}{\tau}\right) + \right.\nonumber\\
 &+&\left.\tau^2\omega_{nn''}^2
 (\tau^2\omega_{nn''}^2 + 1)C_{+}\left(\frac{2t}{\tau}\right) + 15\tau^2\omega_{nn''}^2 -16i\tau\omega_{nn''}
 S_{+}\left(\frac{t}{\tau}\right) + \right.\nonumber\\
 &+&\left.\left.
 2i\tau\omega_{nn''}S_{+}\left(\frac{2t}{\tau}\right)+12\right)-12
 \right]  \;, \label{ngromw5a}\\
  h({t})&=&h_{\kappa_5}({t}) =
h_{\pm,\kappa_5}\left(\frac{t}{\tau}\right) = \frac{i\kappa_5\tau^2}{2\omega_{nn''}(\tau^4\omega_{nn''}^4\pm 5\tau^2\omega_{nn''}^2+4)}
\left[{\rm e}^{it\omega_{nn''}}\left(\mp \tau^4\omega_{nn''}^4 +\right. \right.\nonumber\\
&+&\left.
2i\tau^3\omega_{nn''}^3 S_{\pm}\left(\frac{t}{\tau}\right)\pm 2\tau^2\omega_{nn''}^2
(\tau^2\omega_{nn''}^2\pm 4)C_{\pm}\left(\frac{t}{\tau}\right)\mp \tau^2\omega_{nn''}^2(\tau^2\omega_{nn''}^2 \pm 1)\cdot
\right.\nonumber\\
&\cdot&\left.
 C_{\pm}\left(\frac{2t}{\tau}\right) - 5\tau^2\omega_{nn''}^2\pm 8i\tau\omega_{nn''}S_{\pm}\left(\frac{t}{\tau}\right)
 \mp 2i\tau\omega_{nn''}S_{\pm}\left(\frac{2t}{\tau}\right)- 2i\tau^3\omega_{nn''}^3 \cdot  \right. \nonumber\\
 &\cdot&\left.
 S_{\pm}\left(\frac{2t}{\tau}\right)\mp \left.4 \right) -
 2\tau^2\omega_{nn''}^2\pm 4\right]\;,
\label{ngromw6}
\end{eqnarray}
\begin{eqnarray}
h({t})&=&h_{\kappa_6}({t}) =
h_{\pm,\kappa_6}\left(\frac{t}{\tau}\right) =
\pm\frac{\kappa_6\tau^4}{2(\tau^4 \omega_{nn''}^4\pm 5\tau^2\omega_{nn''}^2 +4)}\left[{\rm e}^{i\omega_{nn''}t}\left( -2(\tau^2\omega_{nn''}^2\pm 4)\cdot\right.\right.\nonumber\\
&\cdot&\left.\left.
C_{\pm}\left(\frac{t}{\tau}\right) +2 (\tau^2\omega_{nn''}^2\pm 1)
C_{\pm}\left(\frac{2t}{\tau}\right) -2i\tau\omega_{nn''}S_{\pm}\left(\frac{t}{\tau}\right)
\left((\tau^2\omega_{nn''}^2\pm 1)\cdot\right.\right.\right.\nonumber\\
&\cdot&\left.\left.\left.
C_{\pm}\left(\frac{t}{\tau}\right)-\tau^2\omega_{nn''}^2\mp 4\right)\right)\pm 6\right]
\;, \label{ngromw7}
\end{eqnarray}
for $n\neq n''$. \\
Consequently, the solution of Schr\"{o}dinger equation at first level of iteration procedure takes the form
\begin{equation}
\psi(t,\bar{x}) = \sum_{n=0}^{\infty}\sum_{l=0}^{n-1}\sum_{m=-l}^{l}c_{nlm}^{(1)}(t) {\rm e}^{iE_nt}\psi_{nlm}(\bar{x}) \;.\label{grom11}
\end{equation}

Finally, it should be noted that  coefficients $c_{nlm}(t)$ have quite simple physical interpretation, i.e. their module is proportional to the probability of
transition from initial state $\psi_0(\bar{x})$ to the final wave function  $\psi_{nlm}(\bar{x})$. Particulary, in accordance with formula
 (\ref{grom10a})\footnote{We consider transition associated with two different energy levels.}, for $\psi_0(\bar{x})=\psi_{n''l''m''}(\bar{x})$ and $j=1$, one gets
\begin{eqnarray}
P^{(1)}(t) = \left|c_{nlm}(t)\right|^2 =
\left(\frac{m''Ze^2}{2\hbar^2}\right)^2 h^2(t)
\left|\left(\psi_{nlm}(\bar{x}),\frac{1}{r^3}\psi_{n''l''m''}(\bar{x})\right)\right|^2
\;.\label{grom13}
\end{eqnarray}
 Of course, for cosmological constant $\tau$ approaching infinity we obtain the proper formula associated with quantum spaces (\ref{nw2})-(\ref{nw7}).

\section{Final remarks}

In this article we investigate the hydrogen atom model defined on twisted acceleration-enlarged Newton-Hooke space-times \cite{nh}. We find
the solution of corresponding Schr\"{o}din-\\
ger equation at first order of perturbation series as well as we calculate the probability of transition
between two  different energy-eigenstates. It should be noted that the presented algorithm can be extended to much more complicated atomic systems such as, for example, helium, orto- and para-hydrogen atoms \cite{24}, \cite{24a}. The works in this direction already started and are in
progress.

\section*{Acknowledgments}
The author would like to thank J. Lukierski and Z. Haba
for valuable discussions. This paper has been financially  supported  by Polish
NCN grant No 2011/01/B/ST2/03354.


\begin{thebibliography}{99}
\bibitem{mech}A. Deriglazov, JHEP 0303, 021 (2003)
\bibitem{mechnext}S. Ghosh, Phys. Lett. B 648, 262
(2007)
\bibitem{qmnext}M. Chaichian, M.M. Sheikh-Jabbari, A. Tureanu, Phys.
Rev. Lett. 86, 2716 (2001)
\bibitem{field}M. Chaichian, P. Pre\v{s}najder and  A. Tureanu,
Phys. Rev. Lett. 94, 151602 (2005)
\bibitem{fieldnext}G. Fiore, J. Wess, Phys. Rev. D
75, 105022 (2007)
\bibitem{snyder}H.S. Snyder, Phys. Rev. 72, 68 (1947)
\bibitem{grav1}S. Doplicher, K. Fredenhagen, J.E. Roberts, Phys. Lett. B 331, 39
(1994)
\bibitem{string1}A. Connes, M.R. Douglas, A. Schwarz, JHEP 9802, 003
(1998)
\bibitem{1a}S. Coleman, S.L. Glashow, Phys. Rev. D 59, 116008
(1999)
\bibitem{1anext}
R.J. Protheore, H. Meyer, Phys. Lett. B 493, 1 (2000)
\bibitem{class1}S. Zakrzewski
; q-alg/9602001
\bibitem{class2}
Y. Brihaye, E. Kowalczyk, P. Maslanka
; math/0006167
\bibitem{oeckl}R. Oeckl, J. Math. Phys. 40, 3588 (1999)
\bibitem{chi}M. Chaichian, P.P. Kulish, K. Nashijima, A. Tureanu, Phys. Lett. B
604, 98 (2004)
\bibitem{dasz1}M. Daszkiewicz,
Mod. Phys. Lett. A 23, 505 (2008)
\bibitem{kappaP}J. Lukierski, A. Nowicki, H. Ruegg and V.N. Tolstoy, Phys. Lett.
B 264, 331 (1991)
\bibitem{kappaG}S. Giller, P. Kosinski, M. Majewski, P. Maslanka
and J. Kunz, Phys. Lett. B 286, 57 (1992)
\bibitem{lie1}
J. Lukierski and M. Woronowicz, Phys. Lett. B 633, 116 (2006)
\bibitem{qdef}O. Ogievetsky, W.B.  Schmidke, J. Wess, B. Zumino, Comm. Math. Phys.
150, 495 (1992)
\bibitem{paolo}
P. Aschieri, L. Castellani, A.M. Scarfone, Eur. Phys. J. C 7, 159
(1999)
\bibitem{nh}M. Daszkiewicz, Acta Phys. Pol. B 41, 1889 (2010)
\bibitem{nh1}M. Daszkiewicz, Mod. Phys. Lett. A 24, 1325 (2009)
\bibitem{bloch}C. Blohmann, J. Math. Phys. 44, 4736 (2003)
\bibitem{wess}J. Wess
; hep-th/0408080
\bibitem{romero}J.M. Romero and J.D. Vergara, Mod. Phys. Lett. A 18,
1673 (2003)
\bibitem{romero1}
J.M. Romero, J.A. Santiago, J.D. Vergara, Phys. Lett. A 310, 9
(2003)
\bibitem{cytowania}Y. Miao, X. Wang, S. Yu; arXiv: 0911.5227 [math-ph]
\bibitem{oscylator}M. Daszkiewicz, C.J. Walczyk, Acta Phys. Pol. B 40, 293
(2009)
\bibitem{daszwal}M. Daszkiewicz, C.J. Walczyk, Phys. Rev. D 77, 105008 (2008)
\bibitem{toporzelek} E. Harikumar, A.K. Kapoor; arXiv: 1003.4603 [hep-th]
\bibitem{kijanka}A. Kijanka, P. Kosinski, Phys. Rev. D 70, 12702
(2004)
\bibitem{giri}P.R. Giri, P. Roy; 0803.4090 [hep-th]
\bibitem{ggggg}J. Lukierski, P.C. Stichel, W.J. Zakrzewski, Annals of Phys. 260, 224 (1997)
\bibitem{pioneer}J.D. Anderson, P.A. Laing, E.L. Lau, A.S. Lin, M.M.
Nieto, S.G. Turyshev, Phys. Rev. Lett. 81, 2858 (1998)
\bibitem{24} L. J. Schiff, {\em Quantum Mechanics}\/, McGraw-Hill, New York 1968
\bibitem{24a}S. Kryszewski, {\em Quantum Mechanics}\/, University of Gdansk, Gdansk 2010
\end{thebibliography}
\end{document}